\documentclass[11pt]{article}
\usepackage{moriond,epsfig}

\def\Journal#1#2#3#4{{#1} {\bf #2}, #3 (#4)}

\def\be{\begin{equation}}
\def\ee{\end{equation}}
\def\bea{\begin{eqnarray}}
\def\eea{\end{eqnarray}}

\begin{document}
\vspace*{0.9cm} 
\title{BLAZARS: THE GAMMA-RAY VIEW OF AGILE}

\author{F. D'Ammando$^{1,2}$, S. Vercellone$^{3}$, I. Donnarumma$^{1}$, A. Bulgarelli$^{4}$,
  A. W. Chen$^{5,6}$, A. Giuliani$^{5}$, F. Longo$^{7}$, L. Pacciani$^{1}$ , G. Pucella$^{1}$,
  M. Tavani$^{1,2}$ and V. Vittorini$^{1,6}$ (on
  behalf of the AGILE Team) \\ } 

\address{$^{1}$INAF-IASF Roma, Via Fosso del Cavaliere 100, 00133 Roma, Italy \\
$^{2}$Dip. di Fisica, Univ. ``Tor Vergata'', Via della Ricerca Scientifica 1, 00133
Roma, Italy \\
$^{3}$INAF-IASF Palermo, Via Ugo La Malfa 153, 90146 Palermo, Italy \\
$^{4}$INAF-IASF Bologna, Via Gobetti 101, 40129 Bologna, Italy \\
$^{5}$INAF-IASF Milano, Via E. Bassini 15, 20133 Milano, Italy \\
$^{6}$CIFS-Torino, Viale Settimio Severo 3, 10133 Torino, Italy \\
$^{7}$Dip. di Fisica and INFN, Via Valerio 2, 34127 Trieste, Italy}

\maketitle\abstracts{Since its launch in April 2007, the AGILE satellite
  detected with its Gamma-Ray Imaging Detector (GRID) several blazars at high
  significance: 3C 279, 3C 454.3, PKS 1510--089, S5 0716$+$714, 3C 273, W Comae,
  Mrk 421 and PKS 0537--441. Moreover, AGILE was able both to rapidly respond
  to sudden changes in blazar activity state at other wavelengths and to alert
  other telescopes quickly in response to changes in the gamma-ray
  fluxes. Thus, we were able to obtain multiwavelength data from other
  observatories such as $Spitzer$, $Swift$, RXTE, $Suzaku$, INTEGRAL, MAGIC,
  VERITAS, as well as radio-to-optical coverage by means of the GASP Project
  of the WEBT and REM. This large multifrequency coverage gave us the
  opportunity to study the Spectral Energy Distribution of these sources from
  radio to gamma-rays energy bands and to investigate the different mechanisms
  responsible for their emission. We present an overview of the AGILE results
  on these gamma-ray blazars and the relative multifrequency data.}

\section{Introduction}

Blazars are a subclass of Active Galactic Nuclei (AGN) characterized by the
emission of strong non-thermal radiation across the electromagnetic spectrum,
from radio to TeV energy bands. The typical observational properties include
irregular, rapid and often very large variability, apparent super-luminal
motion, flat radio spectrum, high and variable polarization at radio and
optical frequencies. These features are interpreted as the result of the
emission of electromagnetic radiation from a relativistic jet that is viewed
closely aligned to the line of sight (Blandford $\&$ Rees \cite{BR}, Urry $\&$
Padovani \cite{UP}). The EGRET instrument onboard Compton Gamma-Ray Observatory
(CGRO) detected for the first time strong and variable high energy $\gamma$-ray emission from
blazars in the MeV--GeV region and together with coordinated multwavelength
observations provided evidence that the Spectral Energy Distributions (SEDs) of
the blazars are typically double humped with the first peak occurring in the
IR/optical band in the Flat Spectrum Radio Quasars (FSRQs) and in UV/X-rays in the BL
Lacertae objects, depending by the total jet power of the source. This first peak is
interpreted as synchrotron radiation from high-energy electrons in a
relativistic jet. The SED second component, peaking at MeV-GeV energies in the FSRQs
and at GeV-TeV energies in the BL Lacs, is commonly interpreted as inverse Compton
scattering of seed photons, internal or external to the jet, by highly relativistic
electrons (Ulrich et al. \cite{Ul}), although other models involving hadronic
processes have been proposed (see e.g. B\"ottcher \cite{Bo} for a recent review). 

With the advent of the AGILE and Fermi-GST $\gamma$-ray satellites, together with
the ground based Imaging Atmospheric Cherenkov Telescopes H.E.S.S., MAGIC and VERITAS, a new
exiting era for the gamma-ray extragalactic astronomy and in particular for
the study of blazars is now open and in
conjunction with a complete multiwavelength coverage will allow us to shed
light on the structure of the inner jet and the emission mechanisms of this class of objects.  

\section{Blazars and AGILE}

\begin{table*}
  \caption{List of the AGILE flaring blazars. 
    References: 1.  Chen et al., 2008, A\&A, 489, L37;
                2.  Giommi et al., 2008, A\&A, 487, L49;
		3.  Donnarumma et al., 2009, ApJL, 691, 13;
                4.  Maier et al., 2009, in preparation;
                5.  Pucella et al., 2008, A\&A, 491, L21;
                6.  D'Ammando et al., 2009, in preparation;
                7.  Pucella et al., 2009, in preparation;
                8.  Pacciani et al., 2009, A\&A, 494, 49;
                9.  Giuliani et al., 2009, A\&A, 494, 509;
                10.  Vercellone et al., 2008, ApJL, 676, 13;
                11. Wehrle et al., 2009, in preparation;
                12. Vercellone et al., 2009a, ApJ, 690, 1018;
                13. Donnarumma et al., 2009, in preparation;
                14. Vercellone et al., 2009b, in preparation;
                15. TBD.
}
\begin{center}  
\begin{tabular}{|l|c|c|c|l|}
\noalign{\smallskip}
    \hline
 \bf{Name} & \bf{Period} & \bf{Sigma} & \bf{ATel $\#$} & \bf {Ref.} \\
 &  {\it start : stop} & & & \\
    \hline
    S5 0716$+$714 & 2007-09-04 : 2007-09-23 & 9.6  & 1221 & 1\\
                  & 2007-10-24 : 2007-11-01 & 6.0  &  -   & 2\\
    Mrk 421
         & 2008-06-09 : 2008-06-15 & 4.5  & 1574, 1583 & 3\\
    W Comae
         & 2008-06-09 : 2008-06-15 & 4.0  & 1582 & 4\\
    PKS 1510$-$089
         & 2007-08-23 : 2007-09-01 & 5.6  & 1199 & 5\\
         & 2008-03-18 : 2008-03-20 & 7.0  & 1436 & 6\\
         & 2009-03-01 : 2009-03-31 & 19.9  & 1957, 1968, 1976 & 7\\
    3C 273
         & 2007-12-16 : 2008-01-08 & 4.6  & -    & 8\\ 
    3C 279
         & 2007-07-09 : 2007-07-13 & 11.1 & -    & 9\\
    3C 454.3
         & 2007-07-24 : 2007-07-30 & 13.8 & 1160, 1167 & 10, 11\\
	 & 2007-11-10 : 2007-12-01 & 19.0 & 1278, 1300 & 12\\
	 & 2007-12-01 : 2007-12-16 & 21.3    & - & 13\\
	 & 2008-05-10 : 2008-06-30 & 15.0    & 1545, 1581, 1592 & 14\\
	 & 2008-07-25 : 2008-08-15 & 12.1    & 1634 & 15\\
    \hline
  \end{tabular}
\end{center}
  \label{tab:blazar_sample}
\end{table*}

AGILE ({\it Astrorivelatore Gamma ad Immagini LEggero}) is an Italian Space Agency (ASI) mission successfully launched
on 23 April 2007 and capable of observing cosmic sources simultaneously in
X-ray and $\gamma$-ray energy bands. The Gamma-Ray Imaging Detector (GRID) constists of a Silicon
Tracker, a non-imaging CsI Mini-Calorimeter and a segmented
anticoincidence system; the GRID is optimized for $\gamma$-ray imaging in the 30
MeV--30 GeV energy band. A co-aligned coded-mask hard
X-ray imager (SuperAGILE) ensures coverage in the 18--60 keV energy band. 

Gamma-ray observations of blazars are a key scientific project of the AGILE
satellite (Tavani et al.\cite{Ta}). In the last two years, the AGILE satellite
detected several blazars during high $\gamma$-ray activity and
extensive multiwavelength campaigns were organized for many of them. 
Table 1 shows the list of AGILE flaring blazars observed up now.
The $\gamma$-ray activity timescales goes from a few days (e.g. S5
0716$+$714) to several weeks (e.g. 3C 454.3 and PKS 1510--089) and the flux
variability observed has been negligible (e.g. 3C 279), very rapid (e.g. PKS
1510--089) or extremely high (e.g. 3C 454.3 and PKS 1510--089). Only
few objects were detected more than once in flaring state by AGILE and only already known $\gamma$-ray emitting
source showed flaring activity. This evidence together with the early results
from the first three months of Fermi-LAT $\gamma$-ray all-sky survey (Abdo et al.\cite{Ab}) suggest possible constraint on the properties of the most intense
$\gamma$-ray emitters. 
In the following section we will present the most interesting results on multiwavelength
observations of the individual sources detected by AGILE.

\section{Individual Sources}

\subsection{3C 454.3}

3C 454.3 is the blazar which exhibited the most variable activity in the $\gamma$-ray
sky in the last two years. In the period July 2007--January 2009 the AGILE
satellite monitored intensively 3C 454.3 together with $Spitzer$, WEBT, REM, MITSuME,
$Swift$, RXTE, $Suzaku$ and INTEGRAL observatories,
yielding the longest multiwavelength coverage of this $\gamma$-ray quasar so
far. The source underwent an unprecedented long period of very high
$\gamma$-ray activity, showing flux levels variable on short
timescales of 24--48 hours and reaching on daily timescale a $\gamma$-ray flux higher than
400$\times$10$^{-8}$ photons cm$^{-2}$ s$^{-1}$. Also the optical flux appears
extremely variable with a brightening of several tenths of magnitude in a few
hours. The comparison of the light curves shows that the emission in the optical and $\gamma$-ray
bands appears to be well correlated, with a time lag less than one day, as
confirmed also by the analysis of the early Fermi-LAT data in Bonning et al.\cite{Bn}. The dominant emission mechanism over 100 MeV
seems to be the inverse Compton scattering of relativistic electrons in the jet on the
external photons from the Broad Line Region (BLR), even if the $\gamma$-ray
spectrum observed by AGILE in December 2007 seems to
require also the contribution of external Compton of seed photons from a hot corona.

\subsection{PKS 1510--089}

PKS 1510--089 showed in the last two years high variability over all the electromagnetic spectrum,
in particular an high $\gamma$-ray activity was detected by AGILE with two
intense flaring episodes in August 2007 and March 2008 and an extraordinary
actitivity during the entire March 2009 with several flaring episode and a
flux reaching 500$\times$10$^{-8}$ ph cm$^{-2}$ s$^{-1}$. The
multiwavelength data carried out by GASP-WEBT and $Swift$ in 2008--2009 seems to indicate
the presence in the spectrum of thermal features quasar-like such as the little blue bump and
the big blue bump. Instead the X-ray spectrum exhibits a soft X-ray excess, of
which the nature is unclear but that could be a feature of the bulk
Comptonization mechanism. Moreover, the $Swift$/XRT observations seems to show a
redder-when-brighter behaviour in X-rays (i.e. the spectrum is harder when the source is
brighter) already observed by Kataoka et al.\cite{Ka} in this source. 

\subsection{3C 279}

3C 279 is the first extragalactic source detected by AGILE in the $\gamma$-ray
band. The average
$\gamma$-ray flux over 4 days of observation is F$_{E>100 MeV}$ = (210 $\pm$ 38)$\times$10$^{-8}$ ph cm$^{-2}$ s$^{-1}$, similar of the high state
observed by EGRET. A strong minimum in the optical band was detected by REM two months before the GRID observations.
The spectrum of this source during the flaring episode observed by
AGILE is soft ($\Gamma$ = 2.22 $\pm$ 0.23) and this could be an
indication of a low accretion state of the disk occurred some
months before the $\gamma$-ray observations, suggesting a dominant
contribution of the external Compton scattering of direct disk (ECD) radiation
compared to the external compton scattering of the Broad Line Region
clouds (ECC). In fact the reduction of the activity of the disk should
cause the decrease of the photon seed population produced by the disk and
then a deficit of the ECC component with respect to the ECD, an effect delayed
of the light travel time required from the inner disk to the BLR. 

\subsection{3C 273}

3C 273 was the first extragalactic source detected simultaneously by the GRID
and SuperAGILE detectors during a multiwavelength campaign over three
weeks between December 2007 and January 2008 involving also simultaneous REM,
$Swift$, RXTE and INTEGRAL coverage. The average flux in the 20--60 energy
band is (23.9 $\pm$ 1.2) mCrab, whereas the source was detected by the GRID only in the second week, with an average flux of F$_{E>100 MeV}$ =
(33 $\pm$ 11)$\times$10$^{-8}$ ph cm$^{-2}$ s$^{-1}$. The comparison of
the light curves seems to indicate a possible anti-correlation between the
$\gamma$-ray emission and the soft and hard X-rays. The SED is consistent with
a leptonic model where the soft X-ray emission is produced by the combination
of SSC and EC models, while the hard X-ray and $\gamma$-ray emission is due to
external Compton scattering by thermal photons of the disk. The spectral variability
between the first and the second week is consistent with the acceleration
episode of the electron population responsible for the synchrotron emission.    

\subsection{S5 0716+714}

The intermediate BL Lac object S5 0716+714 was observed by AGILE during two
different periods: 4--23 September and 23 October -- 1 November 2007. In mid
September the source showed an high $\gamma$-ray activity with an average
flux of F$_{E>100 MeV}$ = (97 $\pm$ 15)$\times$10$^{-8}$ ph cm$^{-2}$ s$^{-1}$
and a peak flux of F$_{E>100 MeV}$ =
(193 $\pm$ 42)$\times$10$^{-8}$ ph cm$^{-2}$ s$^{-1}$. This is one of the most high flux observed by
a BL Lac object. An almost simultaneous GASP-WEBT optical campaign started
after the AGILE detection and the resulting SED is consistent with a
two-components SSC model. Recently Nilsson et al.\cite{Ni} has estimated the
redshift of the source (z = 0.31 $\pm$ 0.08) and this allowed us to calculate the total power
transported in the jet, which results extremely high and at limit of the maximum
power generated by a spinning black hole of 10$^{9}$ M$_\odot$. 

During October 2007, AGILE detected the source at a flux about a factor 2 lower than
September one with no significant variability. Instead, $Swift$ observed strong
variability in soft X-ray, moderate variability at optical/UV and
approximately constant hard X-ray flux. Also this behaviour is compatible with
the presence of two different SSC components in the SED.

\subsection{Mrk 421}

During a ToO towards W Comae on June 2008, AGILE surprisingly
detected also this HBL object. SuperAGILE detected a fast increase of flux
from Mrk 421 up to 40 mCrab in the 15--50 energy band, about a factor 10 higher
than its typical flux in quiescence. The $\gamma$-ray flux detected by GRID, F$_{E>100 MeV}$ =
(42 $\pm$ 13)$\times$10$^{-8}$ ph cm$^{-2}$ s$^{-1}$, is about a factor
3 higher than the average EGRET value, even if consistent with its maximum. An extensive multiwavelength campaign from optical to TeV energy bands
was organized with the participation of WEBT, $Swift$, RXTE, AGILE, MAGIC and
VERITAS. The light curves show a possible correlated variability between
optical, X-rays and the high part of the spectrum. The SED can be interpreted
within the framework of the SSC model in terms of a rapid acceleration of leptons in the jet. A more complex
scenario is that optical and X-ray emission come from different
regions of the jet, with the inner jet region that produces X-rays and is partially
transparent to the optical radiation.



\section*{Acknowledgments}

F. D. warmly thanks the Marie Curie Actions of the European Union for the financial support for the participation to
the Conference. The AGILE Mission is funded by the Italian
Space Agency (ASI) with scientific and progammatic partecipation by the
Italian Institute of Astrophysics (INAF) and the Italian Institute of Nuclear
Physics (INFN).

\end{document}